\documentclass[11pt,twoside]{article}

\usepackage{asp2006}
\usepackage{fancyhdr,graphicx,caption,rotating,multirow,lscape}
\DeclareGraphicsExtensions{.eps,.eps.gz,.ps,.jpg,.tiff}

\begin{document}

\title{DECPHOT: an optimal deconvolution-based photometric reduction method}  

\author{M. Gillon}
\affil{Observatoire de Gen\`eve, 51 Chemin des Maillettes, 1290 Sauverny,
Switzerland }
\author{P. Magain, V. Chantry, G. Letawe, S. Sohy}
\affil{Institut d'Astrophysique et de G\'eophysique,  Universit\'e
  de Li\`ege,  All\'ee du 6 Ao\^ut, 17,  Bat.  B5C, Li\`ege 1, Belgium}
\author{F. Courbin}
\affil{Laboratoire d'Astrophysique, Ecole Polytechnique F\'ed\'erale de Lausanne (EPFL), Observatoire, CH-1290 Sauverny, Switzerland}
\author{F. Pont}
\affil{Observatoire de Gen\`eve, 51 Chemin des Maillettes, 1290 Sauverny,
Switzerland }
\author{C. Moutou}
\affil{LAM, Traverse du Siphon, BP8, Les Trois Lucs, 13376 Marseille Cedex 12,
France }

\begin{abstract} 
A high accuracy photometric reduction method is needed to take full
advantage of the potential of the transit method for the detection and
characterization of exoplanets, especially in deep crowded fields. In this context,  we present DECPHOT, a new deconvolution-based photometry algorithm able to deal with a very high level of crowding and large variations of seeing. It also increases the resolution of astronomical images, an important advantage for the discrimination of false positives in transit photometry. 
\end{abstract}

\section{Introduction}

While the radial velocity method has brought a harvest of $\sim$ 200 extrasolar planets since 1995, the transit method has been presented as the most promising way to detect  earth-like planets around solar-type stars \citep{s1999}, and as sensitive enough to easily detect hundreds of Hot Jupiters using small telescopes operating from the ground (Hornes 2003). Its capacity of detecting earth twins from space is the key assumption of the Kepler mission \citep{bbbb2001}, the most ambitious transit survey planned so far, scheduled for launch in 2008. With this mission and other ambitious surveys such as CoRoT \citep{bbbb1998}, the future of the method seems very bright. Nevertheless, one should worry that the harvest obtained so far from the ground is (very) far from the expected one: only 14 transiting exoplanets are known, from which 3 have been  detected first by radial velocity measurements. 
This discrepancy between great expectations and an interesting but modest harvest is mainly due to previous over-optimistic estimations of the achieved photometric accuracies (see Pont et al., this volume). As outlined by Pont, Zucker \& Queloz \citep{dzq2006}, all transit surveys suffer from the presence of residual correlated noises in their light curves. These 'red' noises are due to the influence of external (atmosphere, detector) and topological (crowding) parameters on the measured photometry. They were not taken into account in previous expected harvest computations, as it was assumed that their influence  could be easily corrected by differential photometry. Recently, clever post-reduction methods dedicated to a better correction of these red noises have been proposed  like Sysrem (Tamuz et al. 2005) and TFA (Kov\'acs et al. 2005). These detrending methods allow indeed to reduce drastically the covariant noises, but not to the level that we can neglect them. These methods are well suited to remove a large fraction of the red noise due to the external parameters, but are less efficient to remove the topological red noise due to the blends. This part of the red noise should in fact be removed efficiently by the reduction method, not by a post-reduction algorithm. In this context, we have developed a new reduction method able to perform optimal photometry even in very crowded fields. 

We present this method called DECPHOT in Sect. 2 and show some efficient applications in Sect. 3. We give our conclusions in Sect. 4.
 
\section{The method}
The MCS algorithm (Magain et al. 1998) is a deconvolution method
specially adapted to astronomical images containing point sources, which
allows to achieve (1) an increase of the angular resolution, (2) a high
accuracy determination of the positions (astrometry) and the intensities
(photometry) of the objects lying on the image. 

Many algorithms have been
proposed to deconvolve images, but generally with rather modest success. It
has been outlined by Magain et al. that the main problem with most of these methods is that
they try to recover the light distribution at full resolution, i.e. they
attempt to perform a total deconvolution. As the observed light distribution
is represented on a pixel grid, with finite pixel size $\Delta x$, the
sampling theorem (Shannon 1949; Press et al. 1989) implies that components of
frequency above the Nyquist frequency $(2\Delta x)^{-1}$ are mixed up with
lower frequency components by aliasing, giving rise to so-called deconvolution
artifacts (e.g.\  Gibbs oscillations) and leading to very poor astrometric and
photometric accuracies. The main principle of the MCS algorithm is thus to perform a partial
deconvolution in order to recover the light distribution at an improved but
finite resolution compatible with the spatial sampling of the resulting image. 

Thus, the total PSF $t(x)$ can be represented as the convolution
of the PSF in the deconvolved image $r(x)$ by a partial PSF
$s(x)$:
\begin{equation}
t(x) = s(x) \ast r(x) 
\end{equation} where $\ast$ stands for the convolution operator. The algorithm thus
performs the deconvolution of the image by the partial PSF $s(x)$
in order to obtain a final PSF $r(x)$ chosen in such a way that
the final result complies to the Shannon theorem (i.e.\  is well sampled).

An algorithm performing this task was presented in Magain et al. (1998), but the determination
of the partial PSF $s(x)$ was not thoroughly addressed. When an
image contains sufficiently isolated point sources, their shape can be used to
determine an accurate PSF. However, this simple PSF determination is not
generally possible in crowded fields, where no star is sufficiently isolated
to provide a suitable measurement of $t(x)$. 

We have thus developed a method allowing to
simultaneously perform a deconvolution and determine an accurate PSF in fields
containing point sources, even if no isolated star can be found \citep{mmmm2006}. It relies on
the minimization of the following merit function: \begin{equation}
S = \sum_{i=1}^N \frac{1}{\sigma_i^2} (d_i - [s \ast f]_i)^2 + \lambda H(s)
\end{equation} where $N$ is the number of pixels within the image, $d_i$ and 
$\sigma_i$ are the measured intensity and standard deviation in pixel $i$, 
$s_i$ is the {\em unknown} value of the PSF and $f_i$ is the intensity of the
deconvolved image in pixel $i$. $H(s)$ is a smoothing constraint on the PSF
which is introduced to regularize the solution and $\lambda$ is a Lagrange
parameter. In the absence of a diffuse background (thus in a field containing only point
sources), the deconvolved light distribution $f$ may be written:
\begin{equation}
f(x) = \sum_{k=1}^M a_k r(x - c_k)
\end{equation} where $M$ is the number of point sources in the image, while
$a_k$ and $c_k$ are free parameters corresponding to the
intensity and position of point source number $k$. Note that the right-hand
side of the equation represents only point sources, so the sky
background is supposed to be removed beforehand.

In Eq. (2), the smoothing constraint on the PSF $H(s)$ is given by:
\begin{equation}
H(s) = \sum_{i=1}^N (s_i -[g \ast s]_i)
\end{equation} where $g$ is a Gaussian function; its width, together with the 
Lagrange parameter $\lambda$, are adjusted in order to obtain a correct smoothing of the partial PSF $s(x)$, i.e. to prevent fitting too high frequencies. 

In order to avoid local minima during the minimization process, the
algorithm proceeds in several steps which are described in Magain et al. (2006). The final result is (1) an accurate PSF and (2) a higher resolution image
containing only point sources, for which intensities and positions are provided by
the algorithm, which thus allows high accuracy astrometry and photometry. 

From this method designed to obtain accurate PSF in crowded fields, we have developed a new method, DECPHOT (DEConvolution PHOTometry), with one major goal in mind: to reach the highest photometric accuracy possible, even in very tricky cases. 

The first improvement brought to  develop DECPHOT concerned the computational time. Indeed, the classical MCS algorithm is very time consuming, and this problem is partially solved in DECPHOT by a linearisation  made possible by the fact that we want to treat hundreds if not thousands of images of the same field \citep{gggg2006}. We treat a reference high resolution high SN image with the standard iterative algorithm, determine the astrometric transformation connecting this reference image to the others, than fix the astrometry and do several cycles composed of matrix inversion (to get the photometry and the sky background) and iterative determination of the PSF (necessary as it is not  an analytical function). 

Another improvement  is the implementation of the
sky background determination into the problem, instead of determining or subtracting it in a previous step. Indeed, fitting a rather smooth surface through seemingly ``empty"
areas may lead to seeing-dependent systematic errors.  A much more robust
method consists in determining the sky background level so that the shape of
all point sources remains the same, irrespective of their intensities. 
A wrong sky level would affect weaker sources much more strongly than
brighter ones.  Forcing all point sources to have the same PSF shape thus
allows an accurate determination of the sky intensity.

To fit the sky background with the other unknowns,  the observed light distribution $d$ is now  written as:
\begin{equation}
d(x) = s(x)\ast \sum_{k=1}^M a_k r(x -
c_k) + b(x)
\end{equation} where the  sky background is represented by the function
$b(x)$, chosen to be relatively smooth. In practice, a second
order polynomial (6 free parameters) has been found suitable for images
obtained in the optical.

For the deconvolution of the reference image, we have now to minimize the
following merit function: \begin{equation}
S = \sum_{i=1}^N \frac{1}{\sigma_i^2} (d_i - b_i-[s \ast f]_i)^2 + \lambda H(s)
\end{equation} 
where $b_i$ is the sky level in pixel $i$, computed through a second-order
polynomial whose coefficients are determined as follows:

\section{Example of transit light curves obtained with DECPHOT}

In the context of the precise determination of the parameters of the OGLE planets, we observed with the VLT and the NTT  telescopes transits of the planets OGLE-TR-10 and OGLE-TR-56 \citep{pppp2006}, OGLE-TR-113 (Gillon et al. 2006) and OGLE-TR-132 (Gillon et al. 2007a). Each photometric reduction was done with DECPHOT. The resulting light curves obtained for OGLE-TR-113$b$ are shown in Fig. 1.

\begin{figure}[!ht]
\centering
\includegraphics[angle=0,width=8cm]{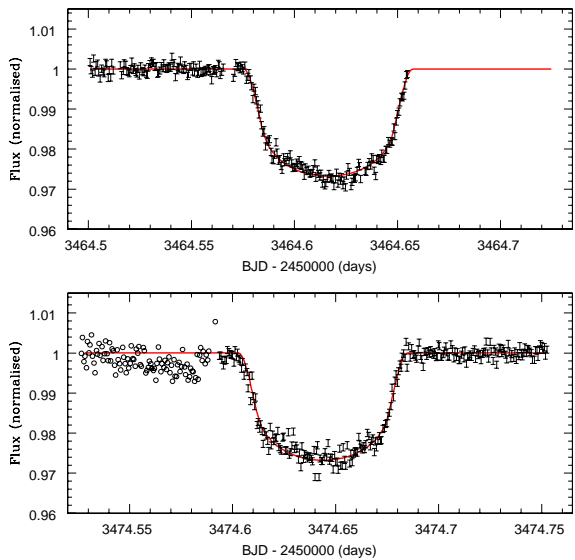}
\caption{NTT/SUSI2 light curve for the first ($top$) and the second ($bottom$) observed transits of
OGLE-TR-113b, with the best-fit transit curve superimposed. For the second transit, the variations of
the flux before the transit are due to a bad column of the CCD
located close to the PSF cores of OGLE-TR-113 and a bright reference star (open
symbols). \label{ogle113}}
\end{figure}

For the first night, the $rms$ of the light curve of OGLE-TR-113 before
the transit is 1.20 mmag, while the mean photon noise is 0.95 mmag. For the
second night, the $rms$ of the light curve after the transit is 1.26
mmag, for the same mean photon noise (0.95 mmag). The slight difference between the photon noise and the observed $rms$ can be explained by  the fact that OGLE-TR-113 has a 0.4 mag brighter visual companion about 3$\arcsec$ to the South (see Fig. 2). 
When a star's PSF is
blended with another one, a part of the noise of the contaminating star
is added to its own noise, resulting in a decrease of the maximal
photometric accuracy attainable. This effect is of course very dependent on the
seeing, and may have a large impact on the final harvest of a transit survey
(Gillon et al. 2007b). The higher seeing during the second transit explains the higher dispersion compared to the first transit. 

\begin{figure}[!ht]
\centering
\includegraphics[angle=0,width=10cm]{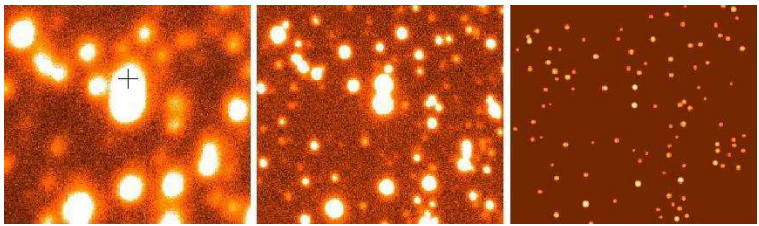}
\caption{OGLE-TR-113 (marked with a cross) in a 256 pixels $\times$ 256 pixels sub-image (0.7 $\arcmin$ $\times$ 0.7 $\arcmin$) from
 the worst (\emph{left}) and best (\emph{middle}) seeing NTT/SUSI2 image of our run ($top$ = North, $left$ = East).
The nearby star just South of OGLE-TR-113 is about 0.4 mag brighter. \emph{Right}: deconvolved image. \label{ogle113_2}}
\end{figure}

\section{Discussion}

Our new photometric reduction method based on the deconvolution of the analyzed images has a high potential for performing optimal photometry in all cases, even in highly crowded fields, as shown by the results obtained in the case of the OGLE planets follow-up. Such an optimal method could increase the potential of the transit surveys, as it appears now clearly that their present potential is still low compared to past expectations, despite the use of sophisticate detrending algorithms. Indeed, the use of DECPHOT should (1) improve the accuracy of the flux measurement, and (2) decrease the topological red noise, a major part of the red noise remaining in the light curves after post-reduction. Furthermore, as DECPHOT allows the user to model the partial PSF with a better sampling than the one of the data, it should give optimal results even in the case of under-sampled data, a frequent case in shallow transit surveys.

The existing version of DECPHOT has a major drawback compared to the other existing reduction method: it is much slower. Extrapolating the processing times obtained in the case of OGLE planets follow-up to transit surveys data, i.e. to thousands of images with a much larger number of pixels ($10^6$ to $10^7$) and sources leads to overwhelming processing times. The problem does not come only from the slowness of the deconvolution of all the images, but also from the preliminary treatment of a reference image to obtain a global solution. This step requires the detection of blended objects undetected in the original image, and this is manually done by the user and should be automatized. We are now working on an improved version of the method which will iterate much faster to the solution and which will incorporate an algorithm able to automatically detect objects hidden in the PSFs in the original image. 

\acknowledgements 
VC is a Research Fellow FNRS (Belgium).

\end{document}